# Unraveling the influence of electronic and magnonic spin current injection near the magnetic ordering transition of IrMn metallic antiferromagnets

O. Gladii,[1,*] L. Frangou,[1] G. Forestier,[1] R. L. Seeger,[1,2] S. Auffret,[1] I. Joumard,[1] M. Rubio-Roy,[1] S. Gambarelli,[3] and V. Baltz[1,**]

[1] *Univ. Grenoble Alpes, CNRS, CEA, Grenoble INP, INAC-Spintec, F-38000 Grenoble, France*
[2] *Departamento de Física, UFSM, Santa Maria, 97105-900 Rio Grande do Sul, Brazil*
[3] *Univ. Grenoble Alpes, CEA, CNRS, INAC-Symmes, F-38000 Grenoble, France*
[*] *olga.gladii@cea.fr*
[**] *vincent.baltz@cea.fr*

**Abstract**

Although spin injection at room temperature in an IrMn metallic antiferromagnet strongly depends on the transport regime, and is more efficient in the case of magnonic transport, in this article, we present experimental data demonstrating that the enhanced efficiency of spin injection caused by spin fluctuations near the ordering temperature can be as efficient for the electronic and magnonic transport regimes. By selecting representative interacting environments, we also demonstrated that the amplification of spin injection near the ordering temperature of the IrMn antiferromagnet is independent of exchange coupling with an adjacent NiFe ferromagnet. In addition, our findings confirm that the spin current carried by magnons penetrates deeper than that transported by conduction electrons in IrMn. Finally, our data indicates that the value of the ordering temperature for the IrMn antiferromagnet is not significantly affected by either the electronic or magnonic nature of the spin current probe, or by exchange coupling.

Antiferromagnetic spintronics explores the spin-dependent transport properties of antiferromagnetically-ordered materials [1–3]. Antiferromagnetic material can be magnetic at the atomic scale and non-magnetic at the macroscopic scale, and as a result has a unique combination of properties: it produces no stray fields and is thus compatible with increased storage densities, it is robust against perturbation due to magnetic fields which is beneficial for data security [4], it can be manipulated and read using mechanisms based on spin-orbit interactions [5–8], and, just as importantly, the writing and propagation of information takes place within picoseconds [9], consequently antiferromagnetic spintronic devices can work hundreds of times faster than their ferromagnetic analogs [10]. The wide range of naturally-occurring antiferromagnet materials - from metals with low to high spin-orbit content to insulators, from collinear to non-collinear and chiral, 3D or 2D spin textures, not to mention the vast array of atomic structures from asymmetric to symmetric, with and without inversion partners - offers a fascinating playground for physicists [1–3]. The many open questions and exciting challenges, combined with a very competitive environment, have led to rapid expansion of this topic over the last decade. Beyond aiming for pure scientific progress, several fields of research have emerged with a view to advancing the development of ultrafast THz components [9], high density secure memories [8,11], artificial neural networks [12], and logic spin current functions and connectors [13,14].

With regards to spin transport in antiferromagnets, several questions have been debated [1]. The efficiency of spin injection was explored through studies of interfacial spin mixing conductance [15–18], a parameter quantifying the amount of spin-angular momentum absorbed at magnetic interfaces upon reflection and transmission. The efficiency of spin propagation was tackled by determining characteristic lengths for the penetration of spins and through studies of various relaxation mechanisms, e.g. dephasing, diffusive, slow relaxation [18–21]. Spin-charge conversion relative to the efficiency of spin-orbit coupling in the core of the antiferromagnet was dealt with by measuring the spin Hall effect and its reciprocal [21–23]. Actually spin transport [24] can be considered to occur by two distinct mechanisms: electronic transport, when spins are carried by conduction electrons; and magnonic transport, which is due to excitation (coherent [25,26] or incoherent [27]) of localized-magnetic-moments. Whereas magnetic insulators only allow magnonic transport, and non-magnetic metals only permit electronic transport, both types of transport regimes can coexist in magnetic metals. Interconversion between the two types of transport occurs at interfaces [28], thus ensuring continuity of the spin flow across heterostructures if the conversion rate is sufficiently efficient. The contribution of electronic and magnonic transport in antiferromagnetic metals is challenging to distinguish, and few results have yet been published on this specific point. Using spin pumping and measuring the inverse spin Hall effect in NiFe/FeMn/W trilayers, Saglam et al. [29] managed to disentangle electronic- (< 2 nm) and magnonic-transport-related (~ 9 nm) penetration depths in FeMn. Other results [1] also appear to suggest that, at room temperature, spin currents propagate more readily when the metallic antiferromagnet is exchange-coupled to a ferromagnet. In this case, magnons produced by the ferromagnet feed directly into the antiferromagnet due to exchange interactions. However, the contribution of interfacial exchange coupling to the initial amplitude of the spin-angular momentum transfer remains controversial. Thus, Tshitoyan et al. [30] demonstrated a direct link between the exchange bias amplitude and the spin-torque efficiency, whereas Saglam et al. [31] reported that spin-orbit torques were independent of the exchange bias direction.

Investigations of the influence of the static vs. fluctuating antiferromagnetic order indicated that spin fluctuations make spin injection more efficient as they open new conduction channels across the interface. As a result, spin injection was shown to be most efficient near the ordering transitions, i.e., near the Néel temperature for an antiferromagnet [32–34].

In this context, in this study we tackled two main questions: whether the magnonic vs. electronic nature of the spin current influences the efficiency of enhanced spin injection near the magnetic phase transition of metallic antiferromagnets; and whether any such enhancement is related to the amplitude of exchange interactions with an adjacent ferromagnet.

Spin currents were generated by the spin-pumping mechanism [35] (Fig. 1(a)). The technique involves inducing resonance in a ferromagnetic spin injector - here a NiFe layer – which is adjacent to a spin sink - here an IrMn layer. We first compared two series of samples consisting of (from substrate to surface) Si/SiO$_2$(500)/NiFe(8)/IrMn($t_{IrMn}$)/Al(2) (nm) multilayers (short name: NiFe/IrMn bilayer) – where mostly magnonic transport is observed, as detailed below – and Si/SiO$_2$(500)/NiFe(8)/Cu(3)/IrMn($t_{IrMn}$)/Al(2) multilayers [33] (short name: NiFe/Cu/IrMn trilayer) – in which mostly electronic transport occurs. It should be noted that data for the NiFe/Cu/IrMn trilayer were adapted from our previous study [33] to make comparison possible. In the NiFe/IrMn bilayers, the IrMn spin-sink can be fed with magnons through direct magnetic coupling with the NiFe spin-injector (Fig. 1(a)). In contrast, in NiFe/Cu/IrMn trilayers, the Cu layer prevents direct magnetic interaction between the IrMn and NiFe layers. The potential magnonic contribution to the spin current in the IrMn layer is therefore the result of electron-magnon conversion mechanisms and is probably less efficient than direct feeding (Fig. 1(b)). We also investigated how spin transport near the ordering transition is influenced by exchange coupling using a series of Si/SiO$_2$(500)/NiFe($t_{NiFe}$)/IrMn($t_{IrMn}$)/Al(2) stacks. For this series, the amplitude of the exchange interactions, specifically of the rotational anisotropy contribution to exchange bias (as explained below), can be tuned by altering the thicknesses of the different layers. $t_{IrMn}$ is the thicknesses of the IrMn layer: $t_{IrMn}$ = 0, 0.6, 0.8, 1 or 1.2 nm; $t_{NiFe}$ is the thicknesses of the NiFe layer: $t_{NiFe}$ = 8, 10, 12, 16, 25, or 50 nm; all thicknesses are given in nanometers. The stacks were deposited at room temperature by dc-magnetron sputtering. The NiFe layer was deposited from a $Ni_{81}Fe_{19}$ (at. %) permalloy target and the IrMn layer was deposited from an $Ir_{20}Mn_{80}$ (at. %) target. An Al cap was deposited to form a protective passivating AlOx film. As part of the NiFe damping enhancement ($\alpha^p$) is a reciprocal effect of spin injection, damping enhancement can be used to investigate spin injection. Spin-pumping experiments (Fig. 1(a,b)) and the corresponding series of ferromagnetic resonance spectra (Fig. 1(c)) were therefore recorded at temperatures (T) ranging between 5 and 300 K, using a continuous-wave electron paramagnetic resonance spectrometer operating at 9.6 GHz and fitted with a cavity. When not specified, the varying bias field was applied in the plane of the sample. For each temperature tested, the peak-to-peak linewidth, $\Delta H_{pp}$, and the resonance field, $H_{res}$, were determined by fitting the NiFe differential resonance spectrum to a Lorentzian derivative (Fig. 1(c)). The total Gilbert damping, $\alpha$, was calculated from $\alpha(T) = \left[\Delta H_{pp}(T) - \Delta H_0(T)\right]\sqrt{3}|\gamma|/(2\omega)$, where $\Delta H_0$ is the inhomogeneous broadening due to spatial variations in the magnetic properties [36], $\gamma$ is the gyromagnetic ratio, and $\omega$ is the angular frequency. The frequency-independent inhomogeneous broadening was determined from frequency-dependent spin-pumping experiments using a separate broadband coplanar waveguide at room temperature (Fig. 1(d)). For all samples, $\Delta H_0$ was one order of magnitude smaller than $\Delta H_{pp}$. We took $\Delta H_0(T) = \Delta H_0(300K)$ since $\Delta H_0$ has been shown to be a temperature-independent parameter [33].

Figures 2(a,b) show $\alpha$ plotted as a function of temperature for series of NiFe/Cu/IrMn trilayers and NiFe/IrMn bilayers with various IrMn spin-sink thicknesses. The reference temperature-dependence of the NiFe Gilbert damping, $\alpha^{ref}(T)$, i.e., in the absence of influence of the IrMn spin-sink, was directly deduced from the measurements performed on the samples with $t_{IrMn} = 0$. $\alpha^{ref}$ can be described as the sum of local intrinsic damping due to intraband and interband scattering [37] and non-local damping mostly associated with the loss of angular

momentum due to spin pumping by an ultra-thin NiFeOx layer. This layer formed naturally at the SiO$_2$/NiFe interface during sputter deposition [38]. The increase of $\alpha^{ref}$ at low temperature was associated with the onset of paramagnetic to antiferromagnetic transition of the NiFeOx layer [38]. Addition of the IrMn layer on top of the NiFe and NiFe/Cu stacks opened another relaxation channel, resulting in an additional contribution to damping, $\alpha^p$. The temperature-dependence of the IrMn contribution to NiFe damping can be directly determined from: $\alpha^p(T) = \alpha(T) - \alpha^{ref}(T)$ (Figs. 2(c,d)). With the NiFe/Cu/IrMn trilayers, the IrMn-thickness-dependence of $\alpha$ and $\alpha^p$ tended to increase at room temperature, with oscillation observed near saturation. This behavior can mostly be related to the finite electronic spin diffusion length (approximately 0.7 nm), as extensively discussed in an earlier work [18]. This phenomenon is beyond the scope of the present paper and will not be further discussed here. For the NiFe/IrMn bilayer, it is impossible to accurately extract the IrMn-thickness-dependence of $\alpha$ and $\alpha^p$ at room temperature since it superimposes on the tail of pronounced peaks in the temperature-dependent data.

From the data presented in Figs. 2(a,b) we observe that all the temperature-dependences of $\alpha$ show a bump. This is because $\alpha^p$ reaches a maximum (Figs 2(c,d)), which itself is the direct consequence of the enhanced dynamical transverse spin susceptibility of IrMn when spins fluctuate near the paramagnetic-to-antiferromagnetic phase transition for the IrMn layer. More precisely, the non-local damping $\alpha^p$ is connected to a quality known as spin mixing conductance, $g^{\uparrow\downarrow}$, as $\alpha^p = (g^{\uparrow\downarrow}/S)|\gamma|\hbar/(4\pi M_{S,NiFe}t_{NiFe})$ [15]. This quality has been presented in a linear-response formalism [32] describing spin pumping near thermal equilibrium, and was found to be linked to the dynamical transverse spin susceptibility of the spin-sink, $\chi_k^R$, through

$g^{\uparrow\downarrow}(T) \propto \sum_k \frac{1}{\Omega_{rf}} \operatorname{Im} \chi_k^R(\Omega_{rf}, T)$, where $k$ is the wave vector, and $\Omega_{rf}$ is the angular frequency of the ferromagnetic spin-injector at resonance. Consequently, the non-local damping is directly connected to the dynamical transverse spin susceptibility of the spin-sink which is enhanced around ordering transitions, i.e., near the critical temperatures ($T_{crit}^{IrMn}$). The results presented here show that spin pumping enhancement near the antiferromagnetic phase transition functions regardless of whether the probe involves spin-wave-like or electronic-like transport. Peak broadening may indicate the formation of short range correlation in the antiferromagnet close to $T_{crit}^{IrMn}$. We note that some early debates suggested that the two-magnon scattering mechanism was at the origin of the bump in temperature-dependence observed for $\alpha$ vs. T. It is now acknowledged that the spin injection enhancement mechanism is at stake, and that two-magnon scattering can be ruled out. More specifically, it was shown for NiFe/CoO bilayers that the position of the bump in $\alpha$ as a function of temperature is frequency-independent and that it corroborates with the ordering transition temperature, which can be measured separately by X-ray magnetic linear dichroism [34]. Similarly, for NiFe/Cu/IrMn trilayers, the bump in $\alpha$ correlated with the ordering transition, measured separately by calorimetry [33,39].

Initially, the amplitude of the enhancement appears to be consistently much smaller in the electronic case (through a Cu spacer) compared to the magnonic one (no Cu spacer) (Figs. 2(a,b)). However, this first impression may be misleading. For example, if we consider $t_{IrMn}$ = 0.6 nm, we have $\left[\alpha^p(300K); \alpha^p(T_{crit}^{IrMn})\right] \square \left[0.2 \times 10^{-3}; 2.9 \times 10^{-3}\right]$ for the NiFe/Cu/IrMn trilayer and $\square \left[2 \times 10^{-3}; 31 \times 10^{-3}\right]$ for the NiFe/IrMn bilayer. Thus, although spin injection in the IrMn layer strongly depends on the transport regime at room temperature - being more efficient in the case of the bilayer ($2 \times 10^{-3}$ vs. $0.2 \times 10^{-3}$) - the spin injection enhancement due to spin fluctuations near the ordering temperature can be equally efficient for both types of transport regimes (here, the enhancement is about 15-fold since $\alpha^p(T_{crit}^{IrMn})/\alpha^p(300K) \square 15$ in both cases). The relative spin injection enhancement, $\delta\alpha^p$, is specified in Fig. 2(c). The plot of the IrMn-thickness-dependence of $\delta\alpha^p$ is shown in Fig. 3(a), showing a clear difference for spin injection enhancement, as $\delta\alpha^p$ is independent of $t_{IrMn}$ in the bilayers but not the trilayers, where it scales as $1/t_{IrMn}$ in line with the predictions proposed by Ohnuma et al. [32]. This result is probably a direct consequence of deeper penetration of the spin current carried by magnons in IrMn compared to that transported by conduction electrons (~0.7 nm, i.e., of the same order as the IrMn thickness in this case, thus explaining the decreased enhancement). This observation further supports the hypothesis that the transport regime is mostly magnonic for the bilayer and electronic for the trilayer. Note that although the penetration of the spin current in the magnonic regime has yet to be reported for IrMn, it seems reasonable to expect similar electronic vs. magnonic behavior to that reported for FeMn [29]: a magnonic spin current propagates over 9 nm whereas its electronic counterpart propagates over less than 2 nm.

The position of the spin pumping maximum can be deduced from Figs. 2(a,b) and Figs. 2(c,d), and the resulting IrMn-thickness-dependence of the ordering temperature is plotted in Fig. 3(b). Data for NiFe/Cu/IrMn trilayers were adapted from our previous study [33], where the position of the spin pumping maximum was initially determined by subtraction of a baseline following the natural trend of the signal. This is equivalent to considering $\alpha$ vs. T (Fig. 2(a)) when determining the maximum and accounting for any slight dispersion in the values of the reference $\alpha^{ref}$, e.g. due to the possible differences in growth reproducibility between samples. However, although reading of the spin pumping maximum may appear clear from $\alpha$ vs. T (Fig. 2(a)), some samples do not show a clear peak in $\alpha^p$ vs. T (Fig. 2(c)), i.e., after subtraction of the same $\alpha^{ref}$ from $\alpha$ for all samples. To further clarify this point, data determined from $\alpha^p$ vs. T (Fig. 2(c)), and considering a constant baseline are also provided in Fig. 3(b). Satisfactory agreement was obtained for all but the thickest sample with the smallest signal amplitude. It should be remembered that the thickness-dependence of the ordering temperature is well described by theoretical models [40,41]. The phenomenological model presented in Zhang and Willis [40] is suitable for use in the thin-layer regime, i.e., when the layer is thinner than the spin-spin correlation length. Here, curve fits using $T_{crit}^{IrMn}(t_{IrMn}) = T_N^{IrMn}(bulk)(t_{IrMn} - d)/(2n_0)$ [40] gave a phenomenological spin-spin correlation length of $n_0 = 2.7$ nm and an interatomic distance of d=0.22 nm for the NiFe/Cu/IrMn trilayer [33]; and of $n_0 = 1.9$ nm and d=0.29 nm for the NiFe/IrMn bilayer. To achieve these fits, we took $T_{N,bulk} = 700$ K [42]. X-ray diffraction measurements performed on similar but thicker (9 nm) samples revealed a (111) growth direction and a related interatomic distance, *d*, of about 0.22 nm, similar to that measured for bulk IrMn [42]. The level of discrepancy observed on $n_0$ between the trilayer and the bilayer may be explained by the fact that IrMn in these samples was grown on different 'buffer' layers (IrMn was grown on a Cu layer in the case of the trilayer whereas it was grown on NiFe in the bilayer). Improvement of the phenomenological spin-spin correlation length (i.e., steeper slope) suggests better growth quality for the bilayers. The small IrMn thicknesses were not compatible with x-ray diffraction experiments to further support this point. However, we note that exchange coupling between the IrMn and NiFe layers cannot be the reason for the improvement in the critical temperature of IrMn with the NiFe/IrMn bilayers compared to the NiFe/Cu/IrMn trilayers. Indeed, an interfacial mechanism of this type would result in a greater enhancement of $T_{crit}^{IrMn}$ for thin layers than for thick ones, which contradicts the results presented in Fig. 3(b). Finally, for $t_{IrMn} = 0.6$ nm, the position of the peak can be seen to be the same for the NiFe/IrMn bilayer and the NiFe/Cu/IrMn trilayer, meaning that this position is not altered by exchange coupling. This observation clearly agrees with the hypothesis that the peak can be used as an indicator of the ordering transition temperature - which is specific to the IrMn antiferromagnet - unlike the exchange bias blocking temperature - which is linked to the interaction between the properties of both the NiFe and the IrMn layers (see below for discussion). We feel it is important to first briefly comment on the temperature-dependence of the resonance field, $H_{res}(T)$. If we return to Figs. 2(e,f), it emerges that for the uncoupled NiFe/Cu/IrMn trilayers the temperature-dependence of the resonance field of the samples containing an IrMn spin-sink is unchanged compared to the reference sample (with no spin-sink), whereas it is significantly altered for the exchange-coupled NiFe/IrMn bilayers. This behavior is known to result from rotational anisotropy [43], i.e., from the presence of uncompensated spins in the IrMn antiferromagnet. These uncompensated spins have a longer relaxation time than the characteristic time for ferromagnetic resonance in the NiFe layer (~10 ns). Due to interfacial coupling, these spins are dragged by the NiFe ferromagnet in a quasi-static experiment (~10 min) but stay still in a dynamic experiment, adding to the anisotropy of the NiFe layer and altering its resonance field. Since interfacial coupling is a temperature-dependent parameter, rotational anisotropy is also temperature-dependent as is the alteration of the resonant field. This situation will be discussed in more detail below. Although damping maxima are observed, the relatively monotonous temperature-dependent behavior of $H_{res}$ for the NiFe/Cu/IrMn samples is a good indication that the process does not involve paramagnetic relaxation [44].

Since there is currently no clear experimental evidence of whether spin transport near the ordering transition of an antiferromagnet is influenced by exchange coupling to a ferromagnet, we further investigated series of Si/SiO$_2$(500)/NiFe($t_{NiFe}$)/IrMn($t_{IrMn}$)/Al(2) stacks for which the amplitude of interfacial coupling between the NiFe and the IrMn layers, and in particular that of the rotational anisotropy contribution is tuned through changes to the thicknesses of the different layers. Figures 4(a-d) show the temperature-dependence of the NiFe layer's Gilbert damping and resonance field, for a range of NiFe ferromagnet thicknesses ($t_{NiFe}$) in two representative series of samples: NiFe($t_{NiFe}$)/IrMn(0.6) and NiFe($t_{NiFe}$)/IrMn(1.2) bilayers (nm). The results confirm that the resonant field is altered due to coupling. The influence of temperature on the resonant field can in fact be described using the modified Kittel

formula [43,45]:

$$\omega = |\gamma|\sqrt{\left(H_{res}(T)+H_{E,st}(T)+H_{rot}(T)\right)\left(H_{res}(T)+H_{E,st}(T)+H_{rot}(T)+4\pi M_s^{eff}\right)}, \quad \text{where}$$

$M_s^{eff}(T) = M_{S,NiFe}(T) - 2K_{S,NiFe}/\left(4\pi M_{S,NiFe}(T)t_{NiFe}\right)$ is the effective magnetization, $M_{S,NiFe}$ is the saturation magnetization (the temperature-dependence of which follows the Bloch equation: $M_{S,NiFe}(T) = M_{S,NiFe}(0)\left(1-\beta T^{3/2}\right)$), $K_S$ is the surface anisotropy, $H_{E,st}$ is the static hysteresis loop shift (static anisotropy contribution due to exchange bias), and $H_{rot}$ is the rotational anisotropy (dynamic anisotropy contribution). The lines in Figs. 4(c,d) clearly show how the values of $H_{res}$ measured differ from the expected values in the absence of coupling. These lines correspond to a fit to the high-temperature data for the NiFe(8)/IrMn(0.6) bilayer (above 100 K, i.e., above the onset of coupling), using the Kittel equation and discarding the exchange bias terms. Data-fitting returned $M_{S,NiFe}(0) = 800$ emu.cm$^{-3}$, $\beta = 1 \times 10^{-5}$ K$^{-3/2}$, and $K_S = 1$ erg.cm$^{-2}$, which are in satisfactory agreement with the expected results for an uncoupled NiFe layer. To extract $H_{rot}(T)$ from $H_{res}(T)$, we recorded hysteresis loops separately at various temperatures (inset in Fig. 5(b)) using a quasi-static vibrating sample magnetometer. The resulting temperature-dependence of the static hysteresis loop shift, $H_{E,st}(T)$, and coercive field, $H_{C,st}(T)$ are shown in Figs. 5(a-d) for the NiFe(t$_{NiFe}$)/IrMn(0.6) and NiFe(t$_{NiFe}$)/IrMn(1.2) bilayer series. As expected, due to rotational anisotropy [43], $H_{E,st}$ starts to increase at a much lower temperature (25 and 75 K for t$_{IrMn}$ = 0.6 and 1.2 nm, respectively) than that at which $H_{res}$ decreases (100 and 250 K for t$_{IrMn}$ = 0.6 and 1.2 nm, respectively, from Fig. 4). The temperature-dependent increase in $H_{C,st}$ is generally thought to be the result of antiferromagnetic grains being dragged by the ferromagnet. These same grains stay still in a dynamic experiment, because they have a longer relaxation time than the characteristic time for ferromagnetic resonance, and consequently contribute to $H_{rot}$. For this reason, the temperature-dependent increase in $H_{C,st}$ usually mirrors the increase in $H_{E,st}$. However, this matching contradicts the present findings, suggesting that other factors also contribute to $H_{rot}$. Figure 6(a) shows the temperature-dependence of $H_{rot}$ deduced from the modified Kittel equation. In general, $H_{rot}$ increases when the NiFe thicknesses is reduced, confirming the interfacial nature of the rotational anisotropy contribution. The temperature-dependence of $H_{rot}$ can in fact be described using the formula: $H_{rot}(T) = J_{int,dyn}(T)/(M_{S,NiFe}(T)t_{NiFe})$, where $J_{int,dyn}$ is the dynamic interfacial exchange constant per unit area. This parameter can be expressed as an effective volume anisotropy, $K_{IrMn,eff}$, as follows: $J_{int,dyn}(T) = K_{IrMn,eff}(T)t_{IrMn}$, with $K_{IrMn,eff}(T) = K^0_{IrMn,eff}(1-T/T_{rot})^\lambda$, in analogy to [46], where $T_{rot}$ is the onset of rotational anisotropy. The temperature-dependence of $H_{rot}$ can therefore be described as follows: $H_{rot}(T) = K^0_{IrMn,eff}t_{IrMn}(1-T/T_{eff})^\lambda/(M_{S,NiFe}(T)t_{NiFe})$. Results of data-fitting using this latter formula are plotted in Fig. 6(a). From this figure, we can conclude that $T_{rot} \sim$ 100 and 300 K for $t_{IrMn}$ = 0.6 and 1.2 nm, respectively, and that these values are independent of $t_{NiFe}$. $M_{S,NiFe}$ was also found to be weakly dependent on $t_{NiFe}$, and remains between 800 and 830 emu.cm$^{-3}$. The temperature-dependence of $H_{rot}$ described above predicts that the plot of $H_{rot}(T)t_{IrMn}/t_{NiFe}$ vs. $T/T_{rot}$ will be universal. Figure 6(b) validates this prediction. However, data for $t_{NiFe}$ = 50 nm depart from the universal behavior, probably as a consequence of the small value of $H_{rot}$ leading to larger errors in its determination. Overall, by averaging over the samples with variable NiFe thicknesses and discarding the values for $t_{NiFe}$ = 50 nm, data-fitting for $H_{rot}(T)$ returned $<\lambda>$ = 1.4 and 1.6; and $<K^0_{IrMn,eff}>$ = 5.8 and 5.9 x 10$^3$ erg.cm$^{-3}$, corresponding to $<J^0_{int,dyn}>$ = (3.5 and 7.1) x $10^{-4}$ erg.cm$^{-2}$ for the series with $t_{IrMn}$ = 0.6 and 1.2 nm, respectively.

Figure 7(a) further shows that the notion of rotational anisotropy can also describe the findings for another measurement configuration, when the dc bias field is applied out of the sample plane, compared to the in-plane configuration previously studied. Data fitting for the out-of-plane configuration (Fig. 7(a)) returned $J^0_{int,dyn}$ = (3.6 and 7.8) x $10^{-4}$ erg.cm$^{-2}$ for the NiFe(8)/IrMn(0.6) and NiFe(8)/IrMn(1.2) bilayers (nm), respectively. These values are in satisfactory agreement with those extracted from in-plane measurements. In Fig. 7(b), we plotted the temperature-dependence of the peak-to-peak linewidth ($\Delta H_{pp}$), which is related to the spin injection efficiency. These data superpose for the in-plane and out-of-plane configurations, a fact that is ascribed to the expected isotropic nature of the dynamic susceptibility for polycrystalline films. We also note that, as mentioned earlier, some early debates suggested that the two-magnon scattering mechanism caused the bump in temperature-dependence observed for $\Delta H_{pp}$ vs. T. However, several experiments now demonstrate that the spin injection enhancement mechanism causes this phenomenon [33,34,39]. The fact that $\Delta H_{pp}$ vs. T superpose for the in-plane and out-of-plane configurations further rules out an influence of two-magnon scattering.

The impact of spin fluctuations on the efficiency of spin pumping in the IrMn antiferromagnet and whether it is influenced by coupling with the NiFe layer can now be discussed by extracting the maximum amplitude of spin pumping, $\alpha^p(T^{IrMn}_{crit})$, for all the NiFe and IrMn thicknesses (see Figs. 4(a,b)). The plot of $\alpha^p(T^{IrMn}_{crit})$ vs. $t_{NiFe}$ for the various IrMn thicknesses is given in Fig. 8(a). To facilitate comparison, Fig. 8(a) also shows the NiFe-thickness-dependence of spin pumping at room temperature, $\alpha^p(300K)$, for $t_{IrMn}$ = 0.6 nm. We note that the NiFe-thickness-dependence of $\alpha^p(300K)$ cannot be accurately extracted for $t_{IrMn}$

> 0.6 nm since it overlaps with the pronounced peaks in the tail of the temperature-dependence. The initial increase of $\alpha^p(300K)$ observed in Fig. 8(a) when the thickness of the NiFe layer is qualitatively reduced agrees with the expected ferromagnetic-thickness-dependence of spin pumping, which in this case should scale as $1/t_{NiFe}$ for Gilbert-like damping - $\alpha^p = (g^{\uparrow\downarrow}/S)|\gamma|\hbar/(4\pi M_{S,NiFe}t_{NiFe})$ [15]. However, fitting the data actually returns a $(1/t_{NiFe})^\gamma$ dependence, with $\gamma = 1.6$. This level of deviation from a pure $1/t_{NiFe}$ dependence observed at room temperature can be explained by additional relaxation processes, such as two-magnon scattering, related to the interface roughness [47]. Most importantly, $\alpha^p(T_{crit}^{IrMn})$ qualitatively shows a similar NiFe-thickness-dependence to $\alpha^p(300K)$, meaning that $\alpha^p(T_{crit}^{IrMn})$ simply reproduces the room-temperature behavior. From this observation we can conclude that spin fluctuations act as a spin injection amplifier - as a consequence of the amplification of $g^{\uparrow\downarrow}$ - and that the amplification factor is independent of the NiFe thickness and thus independent of interfacial coupling. In further support of this conclusion, we note that while the contribution of rotational anisotropy to exchange coupling scales linearly with the thickness of the IrMn layer (see discussion above), Fig. 8(a) shows that $\alpha^p(T_{crit}^{IrMn})$ is virtually independent of $t_{IrMn}$.

Finally, we would like to comment on the NiFe thickness-dependence of $T_{crit}^{IrMn}$ (Fig. 8(b)). As expected from finite size scaling, $T_{crit}^{IrMn}$ scales linearly with the IrMn thickness, for all NiFe thicknesses, i.e., whatever the amplitude of interfacial coupling. We note however that the slope of $T_{crit}^{IrMn}$ vs. $t_{IrMn}$ increases with thicker NiFe 'buffer' layers, suggesting a reduction in the phenomenological spin-spin correlation length, $n_0$, since we recall that $T_{crit}^{IrMn}(t_{IrMn}) = T_N^{IrMn}(bulk)(t_{IrMn} - d)/(2n_0)$ [40]. The plot of $n_0$ vs. $t_{IrMn}$ is shown in the inset in Fig. 7(b). We can once again eliminate exchange coupling between the IrMn antiferromagnet and the NiFe ferromagnet as being the reason for the improvement, because such an interfacial mechanism would result in a more extensive enhancement of $T_{crit}^{IrMn}$ for thin compared to thick IrMn layers, which would contradict our experimental findings. Rather, as in the case of NiFe/IrMn bilayers vs. NiFe/Cu/IrMn trilayers, we infer that such a modification of $n_0$ relates to growth quality and more specifically to better quality growth for IrMn on thick NiFe layers. By reversing the order of the growth of the IrMn and NiFe stacks with $t_{IrMn}$ = 1.2 nm and $t_{NiFe}$ = 25 and 50 nm, we were able to confirm that $T_{crit}^{IrMn}$ can recover the same value as that recorded for growth on thinner NiFe layers.

In conclusion, this paper presents systematic experimental demonstrations of the magnonic vs. electronic nature of the spin current in metallic antiferromagnets, and shows how it influences the efficiency of spin injection enhancement near the magnetic phase transition. The paper also provides information on whether this enhancement relates to the amplitude of interfacial exchange interactions. Spin currents were generated using the spin-pumping mechanism and the systems investigated consisted of uncoupled NiFe/Cu/IrMn trilayers and coupled NiFe/IrMn bilayers, served so as to tune the relative electronic and magnonic transport contributions. Additionally, variable NiFe and IrMn layer thicknesses were used to alter the amplitude of interfacial coupling. Through temperature-dependent ferromagnetic relaxation in thin NiFe films we characterize the efficiency of spin injection and how it was affected by spin fluctuations when scanning the ordering temperatures for the IrMn antiferromagnet. Our results showed that spin injection in IrMn at room temperature strongly depends on the transport regime, and that it is more efficient in the case of magnonic transport. However, we also demonstrated that enhanced spin injection due to spin fluctuations near the ordering temperature can be equally efficient for the two types of transport regimes. In addition, we also found a clear difference in the IrMn thickness dependence of such spin injection enhancement as a direct consequence of deeper penetration of the spin current carried by magnons compared to that transported by conduction electrons. Finally, we observed that spin injection amplification near the IrMn ordering temperature is not influenced by the amplitude of interfacial exchange coupling with the adjacent NiFe layer.

**Acknowledgments**

We acknowledge financial support from ANR [Grant Number ANR-15-CE24-0015-01] and KAUST [Grant Number OSR-2015-CRG4-2626]. We also thank M. Gallagher-Gambarelli for critical reading of the manuscript.

**Figure captions**

Fig. 1. (color online) (a,b) Spin-pumping experiments: out-of-equilibrium magnetization (M) dynamics of the NiFe ferromagnet pumps electronic ($I_S^{el}$) and magnonic spin currents ($I_S^{mag}$). Compared to a NiFe/IrMn bilayer (a), potential transmission of the magnonic spin current in an NiFe/Cu/IrMn trilayer (b) involves additional electron-magnon conversion at interfaces since the non-magnetic Cu only allows electronic transport. (c) Representative series of differential absorption spectra (dχ"/dH *vs*. H) measured at different temperatures (T). The data correspond to a series of measurements for a NiFe(8)/IrMn(1.2) bilayer (nm). The lines were fitted to the data using a Lorentzian derivative. The peak-to-peak linewidth (ΔH$_{pp}$) provides information on the amount of spin current transmitted and absorbed by the IrMn antiferromagnet ($\alpha^p$). (d) Representative frequency-dependence of ΔH$_{pp}$ measured at 300K. The lines are linear fit. The data correspond to measurements for NiFe(8)/IrMn(0.6) and NiFe(8)/IrMn(1.2) bilayers (nm).

Fig. 2. (color online) Temperature (T)-dependence of (a,b) the NiFe layer's Gilbert damping (α), (c,d) the IrMn antiferromagnet contribution to NiFe damping ($\alpha^p = \alpha - \alpha^{ref}$), and (e,f) the NiFe resonance field (H$_{res}$) as a function of the IrMn antiferromagnet's thickness (t$_{IrMn}$) for two representative series of samples: NiFe(8)/Cu(3)/IrMn(t$_{IrMn}$) trilayers and NiFe(8)/IrMn(t$_{IrMn}$) bilayers (nm). (a) and (c) are adapted from our previous work [33] to allow comparison. $\delta\alpha^p$ denotes the extra contribution to damping due to the magnetic phase transition of the IrMn antiferromagnet and $T_{crit}^{IrMn}$ stands for the corresponding critical temperature. In (c), data were shifted vertically to facilitate reading, the native values are (0.2, 1.4, 0.75, and 1.25) x 10$^{-3}$ for t$_{IrMn}$ = 0.6, 0.8, 1, and 1.2 nm.

Fig. 3. (color online) IrMn thickness ($t_{IrMn}$)-dependence of (a) the contribution to damping due to the magnetic phase transition of the IrMn antiferromagnet ($\delta\alpha^p$), and (b) the corresponding critical temperature ($T_{crit}^{IrMn}$) for NiFe(8)/IrMn($t_{IrMn}$) bilayers (nm) and NiFe(8)/Cu(3)/IrMn($t_{IrMn}$) trilayers (determined after subtraction of a baseline in a way that it follows the natural trend of the signal [33], open symbol, or considering a constant baseline, dotted signal). In (a), the dashed line corresponds to a constant fit and the straight line to a linear fit of the data constrained to pass through (0,0). In (b), line fitting was based on the equation presented in Ref. [40] in the thin-layer regime.

Fig. 4. (color online) Temperature (T)-dependence of (a,b) the NiFe layer's Gilbert damping ($\alpha$), and (c,d) the NiFe resonance field ($H_{res}$), for a range of NiFe ferromagnet thicknesses ($t_{NiFe}$), as recorded for two representative series of samples: NiFe($t_{NiFe}$)/IrMn(0.6) and NiFe($t_{NiFe}$)/IrMn(1.2) bilayers (nm). The lines in (c,d) correspond to a fit to the high-temperature data for the NiFe(8)/IrMn(0.6) bilayer, using the Kittel equation and discarding the exchange bias terms.

Fig. 5. (color online) Temperature (T)-dependence of (a,b) the exchange bias coupling field ($H_{E,st}$), and (c,d) the coercive field ($H_{C,st}$) when the NiFe ferromagnet thickness ($t_{NiFe}$) is varied for two representative series of samples: NiFe($t_{NiFe}$)/IrMn(0.6) and NiFe($t_{NiFe}$)/IrMn(1.2) bilayers (nm). The inset in (b) shows representative hysteresis loops at various temperatures with the example of the NiFe(8)/IrMn(1.2) bilayer.

Fig. 6. (color online) Temperature (T)-dependence of (a) the rotational anisotropy ($H_{rot}$) calculated from the data in Figs 4 and 5, for a range of NiFe ferromagnet thicknesses ($t_{NiFe}$), as recorded for two representative series of samples: NiFe($t_{NiFe}$)/IrMn(0.6) and NiFe($t_{NiFe}$)/IrMn(1.2) bilayers (nm). Line fitting is discussed in the text along with the corresponding universal behavior of $H_{rot}t_{NiFe}/t_{IrMn}$ with $T/T_{rot}$ plotted in (b). The line in (b) is a visual guide.

Fig. 7. (color online) Temperature (T)-dependence of (a) the NiFe layer's resonance field ($H_{res}$), and (b) the peak-to-peak linewidth of the NiFe absorption spectrum ($\Delta H_{pp}$), for a bias field applied in- and out-of- the plane of the sample, as recorded for two representative samples: NiFe(8)/IrMn(0.6) and NiFe(8)/IrMn(1.2) bilayers (nm). The straight lines in (a) correspond to a fit to the data, using the Kittel equations and including the exchange bias terms for the low-temperature data. For the sake of comparison, the dashed lines in (a) correspond to a fit for the NiFe(8)/IrMn(0.6) bilayer, discarding the exchange bias terms.

Fig. 8. (color online) (a) NiFe thickness ($t_{NiFe}$)-dependence of (a) the IrMn antiferromagnet contribution to NiFe damping ($\alpha^p$) measured at $T = T_{crit}^{IrMn}$ for NiFe($t_{NiFe}$)/IrMn($t_{IrMn}$) bilayers with $t_{IrMn}$ = 0.6, 0.8, 1 and 1.2 nm, and at T = 300 K when relevant, i.e., for $t_{IrMn}$ = 0.6 nm. The lines are visual guides. (b) Corresponding NiFe thickness-dependence of $T_{crit}^{IrMn}$. Data for inverted IrMn(1.2)/NiFe($t_{NiFe}$) bilayers with $t_{NiFe}$ = 25 and 50 nm are plotted for comparison. Line fitting was based on the equation presented in Ref. [40] in the thin-layer regime and returned a phenomenological spin-spin correlation length, $n_0$. Inset: $n_0$ vs. $t_{IrMn}$.

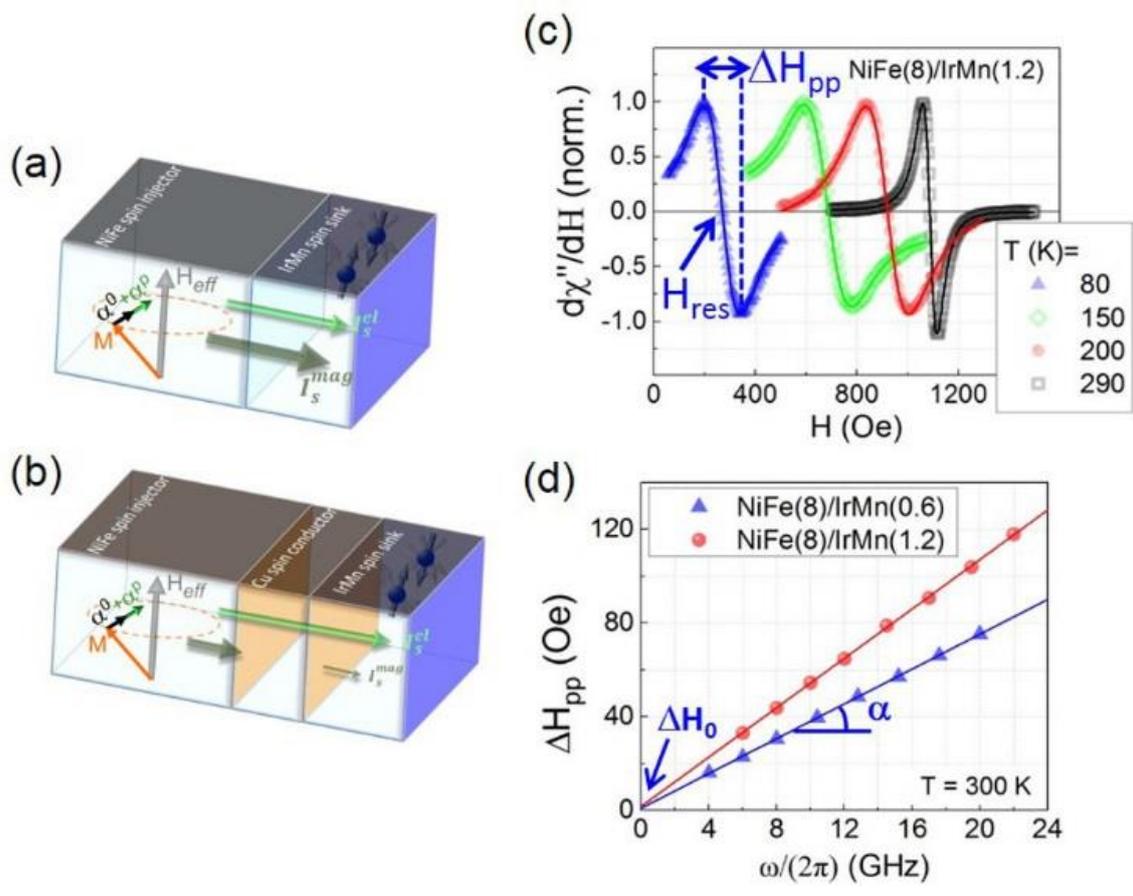

O. Gladii et al                                                                                      Fig. 1

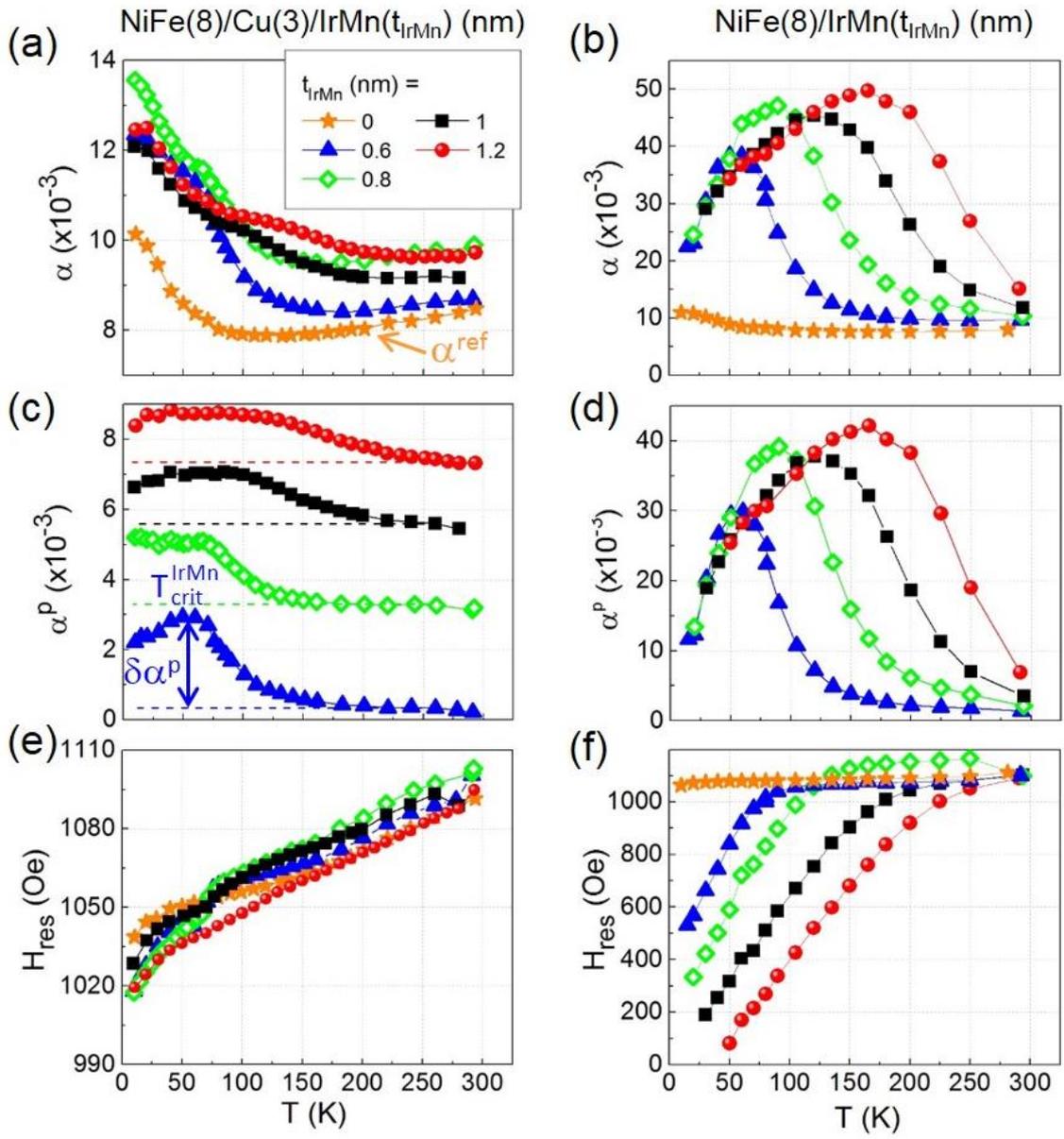

O. Gladii et al                                                                                          Fig. 2

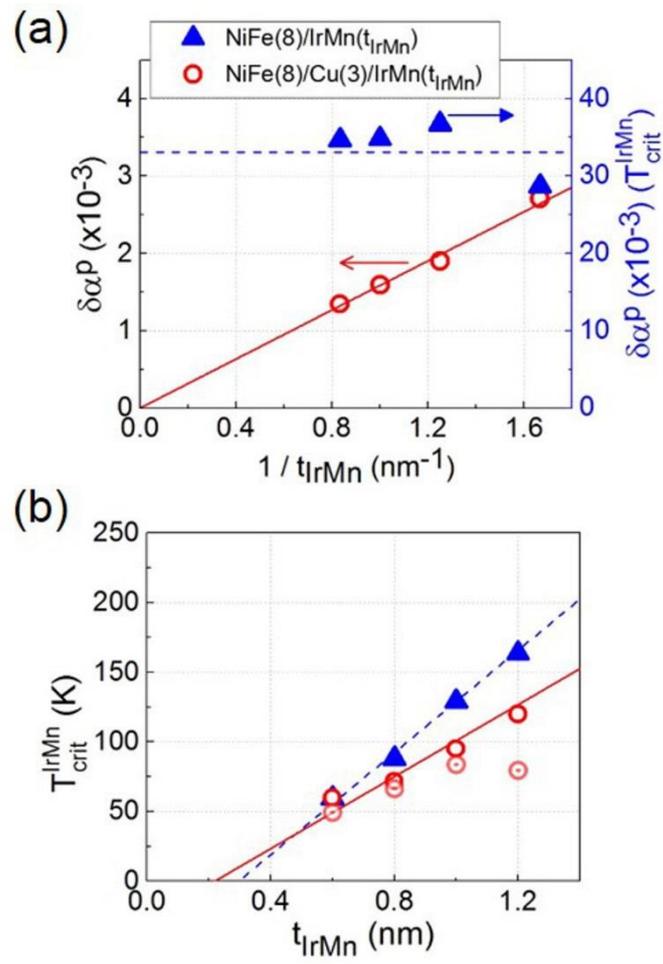

O. Gladii et al            Fig. 3

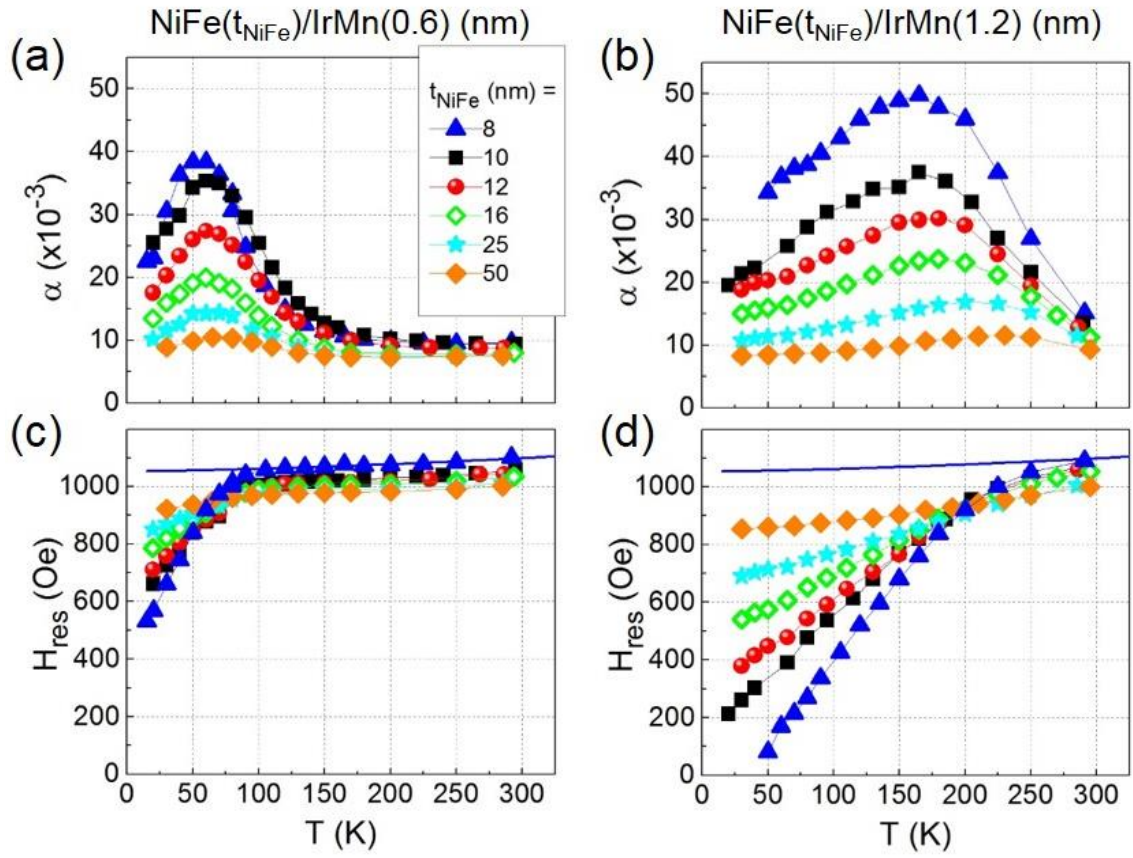

O. Gladii et al Fig. 4

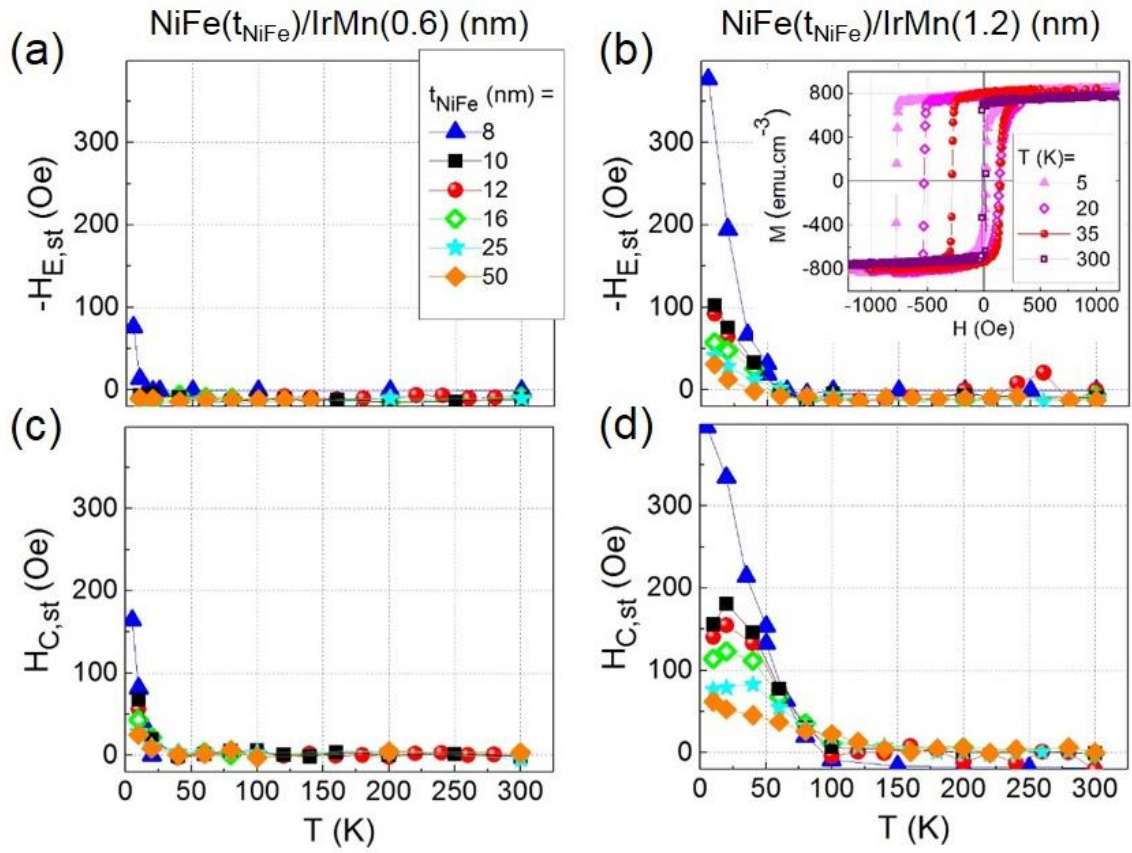

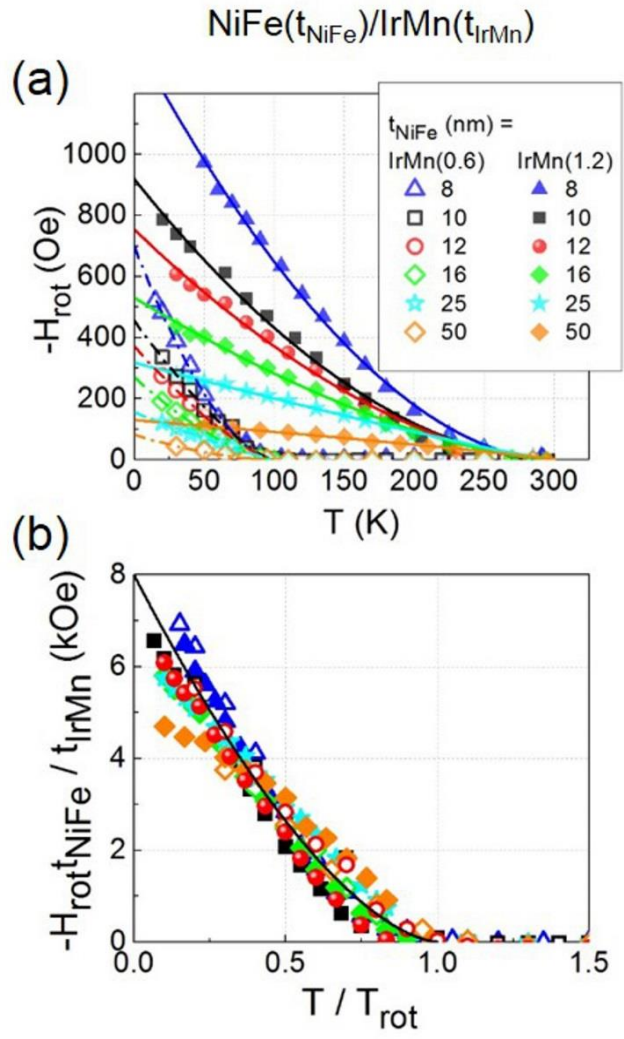

O. Gladii et al    Fig. 6

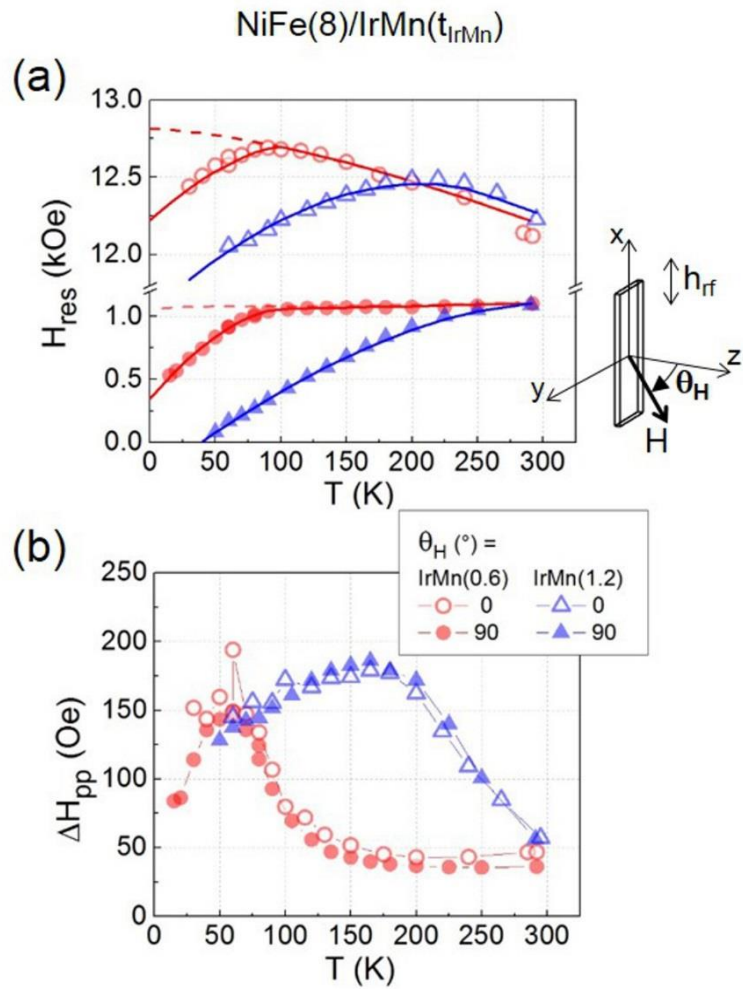

O. Gladii et al · Fig. 7

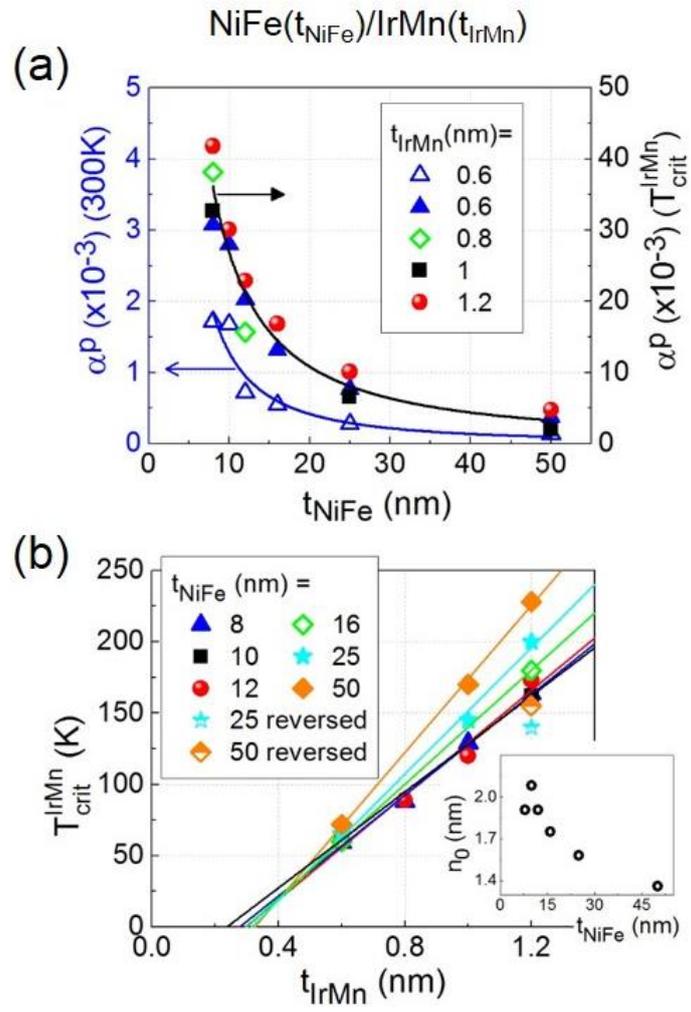